\documentclass[nofootinbib,prd,superscriptaddress,twocolumn]{revtex4}
\usepackage{graphicx}
\usepackage{subfigure}
\usepackage{amsmath, amssymb}
\usepackage{slashed}

\newcommand{\lsim}{\lesssim}

\newcommand{\gsim}{\gtrsim}

\newcommand{\beq}{\begin{equation}}
\newcommand{\eeq}{\end{equation}}
\newcommand{\bea}{\begin{eqnarray}}
\newcommand{\eea}{\end{eqnarray}}
\newcommand{\nn}{\nonumber}

\newcommand{\eps}{\varepsilon}

\newcommand{\mzd}{m_{Z_d}}

\newcommand{\br}{{\rm BR}}
\newcommand{\tev}{{\rm TeV}}
\newcommand{\gev}{{\rm GeV}}
\newcommand{\mev}{{\rm MeV}}
\newcommand{\kev}{{\rm keV}}

\newcommand{\mat}[2][ccccc]{\left( \begin{array}{#1} #2\\ \end{array}\right)}

\begin{document}
\title{\boldmath
Muon $\boldmath g - 2$, Rare Kaon Decays, and Parity Violation from Dark Bosons}
\author{Hooman Davoudiasl}
\affiliation{
Department of Physics, Brookhaven National Laboratory, Upton, New York 11973, USA}
\author{Hye-Sung Lee}
\affiliation{Department of Physics, College of William and Mary, Williamsburg, Virginia 23187, USA}
\affiliation{Theory Center, Jefferson Lab, Newport News, Virginia 23606, USA}
\author{William J. Marciano}
\affiliation{
Department of Physics, Brookhaven National Laboratory, Upton, New York 11973, USA}
\date{February, 2014}

\begin{abstract}
The muon $g_\mu - 2$ discrepancy between theory and experiment may be explained by a light vector boson $Z_d$ that couples to the electromagnetic current via kinetic mixing with the photon.
We illustrate how the existing electron $g_e-2$, pion Dalitz decay,
and other direct production data disfavor that explanation if the $Z_d$ mainly decays into $e^+e^-$, $\mu^+\mu^-$.
Implications of a dominant invisible $Z_d$ decay channel, such as light dark matter, along with the resulting strong bounds from the rare $K \to \pi$ + `missing energy' decay are examined.
The $K$ decay constraints may be relaxed if destructive interference effects due to $Z - Z_d$ mass mixing are included.
In that scenario, we show that accommodating the $g_\mu-2$ data through relaxation of $K$ decay constraints leads to interesting signals
for {\it dark parity violation}.
As an illustration, we examine the alteration of the weak mixing angle running at
low $Q^2$, which can be potentially observable in polarized electron scattering or atomic physics experiments.
\end{abstract}
\maketitle

\section{Introduction}
Dark matter constitutes about $22 \%$ of the Universe's energy-matter budget \cite{Beringer:1900zz}.
However, its exact nature remains elusive.
Various speculative ideas have been proposed based on cosmologically stable candidate particles ranging in mass from below $1 ~\gev$ to above $1 ~\tev$.
Beyond gravity, dark matter interactions and other properties such as spin and extended spectroscopy are also uncertain, with conflicting evidence coming from sensitive underground experiments and astrophysical measurements.

An interesting generic property of some dark matter scenarios is the existence of a broken $U(1)_d$ gauge symmetry in the dark particle sector.  Originally introduced to explain various astrophysics anomalies such as high energy positron excesses or $511 ~\kev$ photons originating from the galactic center \cite{darkPhoton}, it has also been used to provide a novel explanation \cite{LightBosonG-2,Pospelov:2008zw} for the $3.6 \sigma$ discrepancy between the muon's experimental anomalous magnetic moment, $a_\mu \equiv (g_\mu - 2) / 2$, and the Standard Model (SM) prediction.
Employing the $U(1)_d$ gauge symmetry to accommodate this discrepancy is the guiding focus of this paper.
In the simplest scenario, the low mass $U(1)_d$ gauge boson known as the dark photon (or dark $Z$),
$Z_d$, interacts with the SM particles via kinetic mixing with the photon, parametrized by $\eps \ll 1$.

Existing experimental constraints on $\eps$ as a function of $m_{Z_d}$ are reviewed and updated in Sec.~\ref{sec:review}.
There, we discuss the region of parameter space favored by the discrepancy between measured and predicted values
of $a_\mu$ as well as the bound (roughly $m_{Z_d} \gsim 20 ~\mev$)
that follows from a comparison of experiment and theory for the electron anomalous magnetic moment.
Bounds from $\pi^0 \to \gamma Z_d$ searches in Dalitz decays, $\pi^0 \to \gamma e^+ e^-$, and direct $Z_d$ production in electron scattering
are also displayed.

Except for $a_\mu$ and $a_e$, most dark photon constraints assume $\br (Z_d \to e^+ e^-) \simeq 1$ for $m_{Z_d} < 2 m_\mu$ \cite{Essig:2013lka}.
Those bounds can be significantly relaxed if instead light ``dark'' particles exist with masses less than $m_{Z_d} / 2$ and dominate
the branching fractions via $Z_d \to $ invisible decays \cite{Anchordoqui:2012qu,Weinberg:2013kea}.
However, as we describe in Sec.~\ref{sec:kaon}, the decay $K^+ \to \pi^+ + $ missing energy constraints then apply and continue to rule out
large parts of the dark photon parameter space favored by the $g_\mu - 2$ discrepancy.
In particular, the regions around $m_{Z_d} \sim 100$ and $200 ~\mev$ are already severely constrained.

In addition to the $K^\pm \to \pi^\pm Z_d$ loop induced amplitude that arises from kinetic mixing,
an amplitude of potentially similar magnitude can also arise from $Z - Z_d$ mass matrix mixing.
We briefly review the latter formalism in Sec.~\ref{sec:darkZ}.
If destructive interference between the two amplitudes occurs, the $K^\pm \to \pi^\pm Z_d$ bound can be significantly relaxed as shown in Sec.~\ref{sec:kaonrevisit}.
Such a cancellation requires a relationship between $\eps$ and the size of mass matrix mixing parametrized by a small quantity $\eps_Z$.
We describe in Sec.~\ref{sec:WeinbergAngle} how that relation leads to interesting
definite predictions (dark parity violation) that are potentially observable at low $Q^2$ in the running of $\sin^2 \theta_W (Q^2)$, where $\theta_W$ is the weak mixing angle.
Finally, in Sec.~\ref{sec:conclusion}, we present our conclusions.

\section{The Status of Dark Photon Searches}
\label{sec:review}
\begin{figure}[t]
\begin{center}
\includegraphics[width=0.45\textwidth,clip]{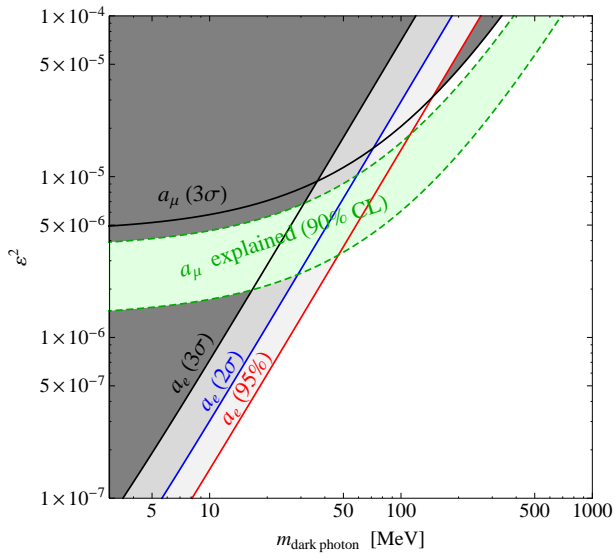}
\end{center}
\caption{Dark photon parameter space with bounds that are independent of the dark photon decay branching ratio.
The green band is the region within which the $3.6 \sigma$ deviation in $a_\mu$ can be explained by the dark photon ($90 \%$ C.L.).
The three $a_e$ curves represent $3\sigma$, $2\sigma$, and $95 \%$ C.L. bounds.
}
\label{fig:al}
\end{figure}

The interaction of a dark photon $Z_d$ corresponding to a broken $U(1)_d$ gauge symmetry in the dark particle sector with the SM is induced by
$U(1)_Y$ and $U(1)_d$ kinetic mixing \cite{Holdom:1985ag} in the Lagrangian
\beq
{\cal L}_\text{gauge} = -\frac{1}{4} B_{\mu\nu} B^{\mu\nu} + \frac{1}{2} \frac{\eps}{\cos\theta_W} B_{\mu\nu} D^{\mu\nu} - \frac{1}{4} D_{\mu\nu} D^{\mu\nu} \label{eq:kinetic}
\eeq
where
\beq
B_{\mu\nu} = \partial_\mu B_\nu - \partial_\nu B_\mu , \quad D_{\mu\nu} = \partial_\mu {Z_d}_\nu - \partial_\nu {Z_d}_\mu ,
\eeq
and $|\eps| \lsim 10^{-2}$ is a (potentially loop induced) mixing parameter.
It can be viewed as an effective counter-term whose value is to be determined experimentally or in some models may be finite and calculable.

After field redefinitions employed to eliminate the cross term in Eq.~\eqref{eq:kinetic}, a coupling of the dark photon to the ordinary electromagnetic current is induced.
\beq
{\cal L}_\text{dark $\gamma$} = -\eps e J^\mu_{em} {Z_d}_\mu , \quad J^\mu_{em} \equiv Q_f \bar f \gamma^\mu f + \cdots . \label{eq:darkPhoton}
\eeq
where $Q_f$ is the electric charge of a given fermion $f$ and the ellipsis represents non-fermionic currents.
At leading order, the effective coupling is basically
given by the $\gamma - Z_d$ mixing parametrized by $\eps$.
Since $|\eps|$ is very small, the next-to-leading order $\eps^2$ as well as ${\cal O}(\eps \, m_{Z_d}^2/m_Z^2)$ effects
can be neglected in the phenomenology we consider in this paper.

An attractive feature of the dark photon model is that there are only 2
parameters in its phenomenological description: dark photon mass ($m_{Z_d}$) and kinetic mixing angle ($\eps$).
The effective coupling of the dark photon to SM particles is the same as that of the photon but
suppressed by $\eps$.

\begin{figure}[t]
\begin{center}
\includegraphics[width=0.45\textwidth,clip]{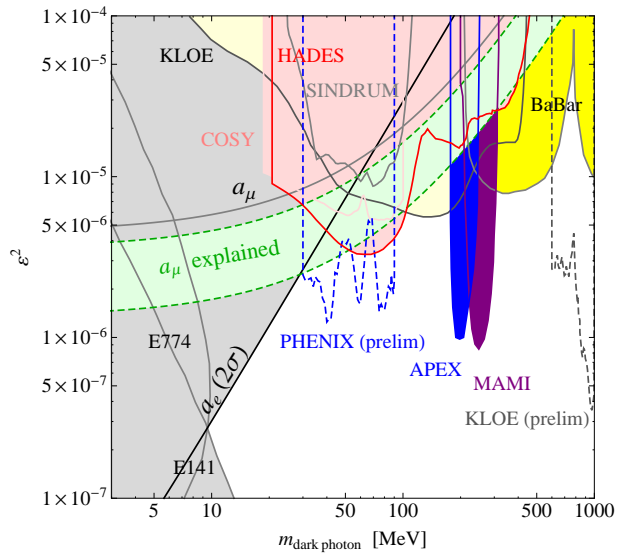}
\end{center}
\caption{Present bounds from various experiments on the dark photon parameter space.
Some of these were obtained with the assumption of $\br (Z_d \to \ell^+ \ell^-) = 1$.
}
\label{fig:currentAndFuture}
\end{figure}

Figure~\ref{fig:al} shows the dark photon parameter space (in the $m_{Z_d} - \eps^2$ plane) along with the constraints from the electron and muon anomalous magnetic moments.
These bounds are more robust compared to most other constraints,
such as those from dilepton bump searches from $Z_d$ decays, in the sense that they
do not depend on the assumed decay branching ratios of the dark photon.

The muon anomalous magnetic moment theory and experiment exhibit a $3.6 \sigma$ discrepancy \cite{Beringer:1900zz}
\beq
\Delta a_\mu = a_\mu^{\rm exp} - a_\mu^{\rm SM} = 288 \, (80) \times 10^{-11}
\eeq
with a slight change in the last digit made from a recent improved QED calculation \cite{Aoyama:2012wj}, Higgs mass of $126 ~\gev$ \cite{ATLAS:2013oma,CMS:ril}, and a small change in the experimental value \cite{Beringer:1900zz}.
The long standing discrepancy could be an early
hint of new physics \cite{Czarnecki:2001pv}, assumed here to be the $Z_d$.

A one loop contribution of the dark photon to the $a_\ell$ ($\ell = e$, $\mu$) \cite{LightBosonG-2,Davoudiasl:2005ks,Pospelov:2008zw,Leveille:1977rc} is given by
\bea
&& a_\ell^{Z_d} = \frac{\alpha}{2\pi} \eps^2 F_V\left(m_{Z_d} / m_\ell \right) \\
&& F_V(x) \equiv \int_0^1 dz \frac{2 z (1-z)^2}{(1-z)^2 + x^2 z} , \quad F_V(0) = 1\,.
\eea
The parameter region of the dark photon that accommodates the $a_\mu$ deviation, using the fine structure constant $\alpha = 1 / 137.036$,
is indicated by the green band (90\% C.L.) in Fig.~\ref{fig:al}.
That figure also contains the $a_\mu$ bound at $3 \sigma$ C.L. and $a_e$ bounds at
$3 \sigma$, $2 \sigma$, and $1.64 \sigma$ (95\% as it is one-sided) C.L. using the constraint
\beq
\Delta a_e = -1.05 \, (0.82) \times 10^{-12}
\eeq
of Ref.~\cite{Aoyama:2012wj}.
The $\Delta a_e$ value was significantly improved recently \cite{Davoudiasl:2012ig,Endo:2012hp,Hanneke:2008tm} by updates in the value of $\alpha$ and improvements in theory \cite{Bouchendira:2010es,Aoyama:2012wj}.
In the subsequent plots, we will employ only the $2 \sigma$ bound on $a_e$.

There are additional bounds on the dark photon parameters from various experiments (Fig.~\ref{fig:currentAndFuture}).
They include beam dump experiments \cite{Andreas:2012mt}, rare meson decays ($\Upsilon$ decays at BaBar \cite{Bjorken:2009mm}, $\phi$ decays at KLOE \cite{Babusci:2012cr}, $\pi^0$ decays at SINDRUM \cite{MeijerDrees:1992kd,Gninenko:2013sr}, WASA-at-COSY \cite{Adlarson:2013eza}, HADES \cite{Agakishiev:2013fwl}, $\eta$ decays at HADES \cite{Agakishiev:2013fwl}), and fixed target experiments (MAMI \cite{Merkel:2011ze}, APEX \cite{Abrahamyan:2011gv}).
There are also preliminary bounds from KLOE 2012 for high mass region ($m_{Z_d} > 600 ~\mev$) \cite{Curciarello:2013lra}
and the PHENIX experiment at BNL RHIC ($\pi^0$ decays) \cite{PHENIXprelim}.

Furthermore, there are ongoing and proposed experiments to test the remaining green band and other parameter space.
CERN NA48/2 experimental data (Dalitz decays of $\pi^0$ from $K^\pm \to \pi^\pm \pi^0$) are under analysis
and their sensitivity can cover $\eps^2 \gsim$ several $\times 10^{-7}$ \cite{Amaryan:2013eja}.
There are direct dark photon searches which use an electron beam with a fixed target of typically high atomic number \cite{Bjorken:2009mm}
to produce $Z_d$s.  The radiated dark photon can decay into a dilepton
forming a resonance over the smooth SM off-shell photon background ($\gamma^* \to e^+ e^-$).
At Mainz, MAMI 2012-2013 experimental data are under analysis \cite{Beranek:2013yqa}.
At Jefferson Lab (JLab) in Virginia, there are 3 proposed or approved
searches for the dark photon (APEX, HPS, DarkLight) \cite{Dudek:2012vr} using fixed target experiments.
Similar searches have also been proposed using the VEPP-3 facility \cite{Wojtsekhowski:2012zq} at the Budker Institute in Russia .
For a recent discussion of the signal and background estimation in the fixed target experiments, see Ref.~\cite{Beranek:2013nqa}.

Figure~\ref{fig:currentAndFuture} shows the currently available bounds on the dark photon parameter space from the
aforementioned experiments for a typical mass range of $m_{Z_d} \approx {\rm few}~\mev - ~\gev$.
(For an overview and overall constraints for wider ranges of parameter space, see Refs.~\cite{Jaeckel:2013ija,Essig:2013lka}.)
Most of these bounds depend on the dark photon decay branching ratios and generally assume
\beq
\br (Z_d \to \ell^+ \ell^-) \equiv \br (e^+ e^-) + \br (\mu^+ \mu^-) = 1
\eeq
for $m_{Z_d} \lsim 300 ~\mev$.
The current published constraints, including a $2 \sigma$ bound from $\Delta a_e$,
only allow a rather tightly constrained parameter region in the green band:
$m_{Z_d} \sim 30 - 50 ~\mev$ and $\eps^2 \sim (2-4)\times 10^{-6}$.
Most of this region will be covered if we include preliminary bounds
from the PHENIX experiment \cite{PHENIXprelim}.
Note that the $\Delta a_e$ bound is expected to improve
with ongoing or planned efforts in the measurement of
both $a_e$ and $\alpha$ \cite{Giudice:2012ms}, which are
independent of the $Z_d$ decay branching ratio.
In short, nearly the entire green band that can explain the $a_\mu$ deviation is
already excluded or under close scrutiny by various experiments, as is clear from Fig.~\ref{fig:currentAndFuture}.

Many of the constraints in Fig.~\ref{fig:currentAndFuture} assume $Z_d$ decays into an observable $\ell^+ \ell^-$ pair with invariant mass $m_{Z_d}$ and branching ratio $\sim 1$.
If, instead, the primary $Z_d$ decay mode is into invisible particles, such as light dark matter pairs with particle masses $< m_{Z_d} / 2$, that effect would negate essentially all the bounds in Fig.~\ref{fig:currentAndFuture}, except those coming from anomalous magnetic moments.
For the case of light dark matter coupled to $Z_d$ with strength ${q_d}_\text{light}$ $g_d$, where ${q_d}_\text{light}$ is its $U(1)_d$ charge and $g_d$ is the gauge coupling, $Z_d \to$ light ``invisible'' matter will be dominant for $| {q_d}_\text{light} \, g_d | \gsim \eps e$, which for the region in $\eps$ we subsequently consider $|\eps| \sim 2 \times 10^{-3}$ suggests (with $\alpha_d = g_d^2 / 4 \pi$)
\beq
3 \times 10^{-8} \lsim {q_d}_\text{light}^2 \alpha_d \lsim 10^{-2} \label{eq:darkBound}
\eeq
as an interesting range for discussion.
The upper bound in that range is somewhat arbitrary, but appropriate for the models we subsequently consider.
We do note, however, that for larger ${q_d}_\text{light}^2 \alpha_d \sim 0.1$ experimental constraints from beam dump experiments \cite{Diamond:2013oda,BES} are already providing interesting bounds.
They effectively assume $Z_d$ boson production $(\sim \eps^2 \alpha)$ followed by $Z_d$ decay and subsequent detection of the $Z_d$ decay products.
So, even for a primary $Z_d \to$ light dark matter scenario, detection is possible if ${q_d}_\text{light} g_d$ is relatively large.
Under those circumstances, they are likely to rule out much of the $\Delta a_\mu$ discrepancy band.
(For examples of future beam dump experiments designed to search for light ``dark'' matter, see Refs.~\cite{Batell:2009di,Dharmapalan:2012xp,Izaguirre:2013uxa,Andreas:2013lya}.)

Instead of explaining constraints from beam dump experiments, which is beyond the scope of this paper, we will next concentrate on the decay $K^\pm \to \pi^\pm Z_d$, $Z_d \to$ light ``dark'' matter which is insensitive to the value of ${q_d}_\text{light}^2 \alpha_d$ as long as it falls in the range of Eq.~\eqref{eq:darkBound}.

\section{Dark Kaon decays}
\label{sec:kaon}
\begin{figure}[t]
\begin{center}
\includegraphics[height=0.27\textwidth,clip]{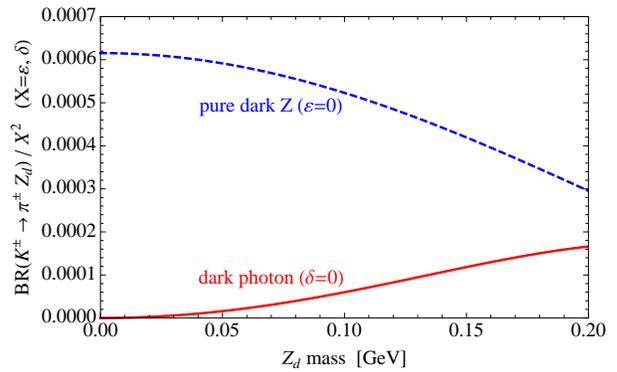}
\end{center}
\caption{$\br (K^\pm \to \pi^\pm Z_d)$ in the dark photon limit ($\delta = 0$, solid line) and the pure dark $Z$ limit ($\eps = 0$, dashed line), as a function of $m_{Z_d}$ for both $\eps$ (dark photon) and $\delta$ (pure dark $Z$).
Here, $m_{H^\pm} = 160 ~\gev$ is assumed.}
\label{fig:BRs}
\end{figure}

As seen in the previous section, flavor conserving meson
decays provide strong constraints on dark photon phenomenology.  We now consider the impact of flavor changing kaon decays on
dark photon parameters.
Earlier discussions about $Z_d$ implications for meson decays in various contexts can be found in the literature \cite{Fayet:2006sp,Kahn:2007ru,Pospelov:2008zw,Batell:2009yf,Barger:2011mt,Davoudiasl:2012ag,Kahn:2012br,Beranek:2012ey}.

From the formalism given in the Appendix, the rate for $K^\pm \to \pi^\pm Z_d$, assuming
a kinetically mixed dark photon, is given by
\bea
&&\Gamma (K^\pm \to \pi^\pm Z_d)|_\eps = \frac{\eps^2 \alpha W^2}{2^{10} \pi^4} \frac{m_{Z_d}^2}{m_K^7}
\sqrt{\lambda (m_K^2, m_\pi^2, m_{Z_d}^2)}\nonumber \\
&&\times \left[(m_K^2 - m_\pi^2)^2 - m_{Z_d}^2(2 m_K^2 + 2 m_\pi^2 - m_{Z_d}^2) \right]
\label{Kdphot}
\eea
which is consistent with the results in Ref.~\cite{Pospelov:2008zw}.  
This process is suppressed for small $m_{Z_d}$ and $m_{Z_d} \simeq m_K - m_\pi$ (the end of phase space).  The branching ratio
associated with the rate in Eq.~(\ref{Kdphot}) is presented in Fig. \ref{fig:BRs} by the
solid curve, as a function of $m_{Z_d}$.

Eq.~\eqref{Kdphot} and the uncertainties of the experimentally measured branching ratios of $K^\pm \to \pi^\pm \ell^+ \ell^-$ \cite{Batley:2009aa,Appel:1999yq,Beringer:1900zz} roughly yield for $Z_d \to \ell^+ \ell^-$
\beq
\eps^2 \lsim \frac{10^{-4}}{\br (Z_d \to \ell^+ \ell^-)} \left( \frac{100 ~\mev}{m_{Z_d}} \right)^2 \,.
\label{eq:KboundsDarkPhoton}
\eeq
This result does not give significant constraints over the existing bounds of $\eps^2 \lsim 10^{-5}$ in the
parameter region of interest
$m_{Z_d} \lsim 300$~MeV (cf. Fig.~\ref{fig:currentAndFuture}), for typically assumed $\br (Z_d \to \ell^+ \ell^-) = 1$.
The situation gets worse if $Z_d$ decays primarily
into very light dark matter or other invisible particles dominantly, lowering $\br (Z_d \to \ell^+ \ell^-)$.

The BNL E949 experiment combined with E787 results \cite{Artamonov:2009sz} measured the illusive $K^+ \to \pi^+ \nu \bar\nu$ and gave upper bounds on the $\br (K^+ \to \pi^+ Z_d)$ as a function of the $Z_d$ mass, 
for $Z_d \to$ `missing energy.'
The region around $m_{ee} \sim 140 ~\mev$ is not constrained, corresponding to
events that were vetoed to avoid the large background from $K^+ \to \pi^+ \pi^0$.

\begin{figure*}[tbh]
\begin{center}
\subfigure[]{
\includegraphics[width=0.45\textwidth,clip]{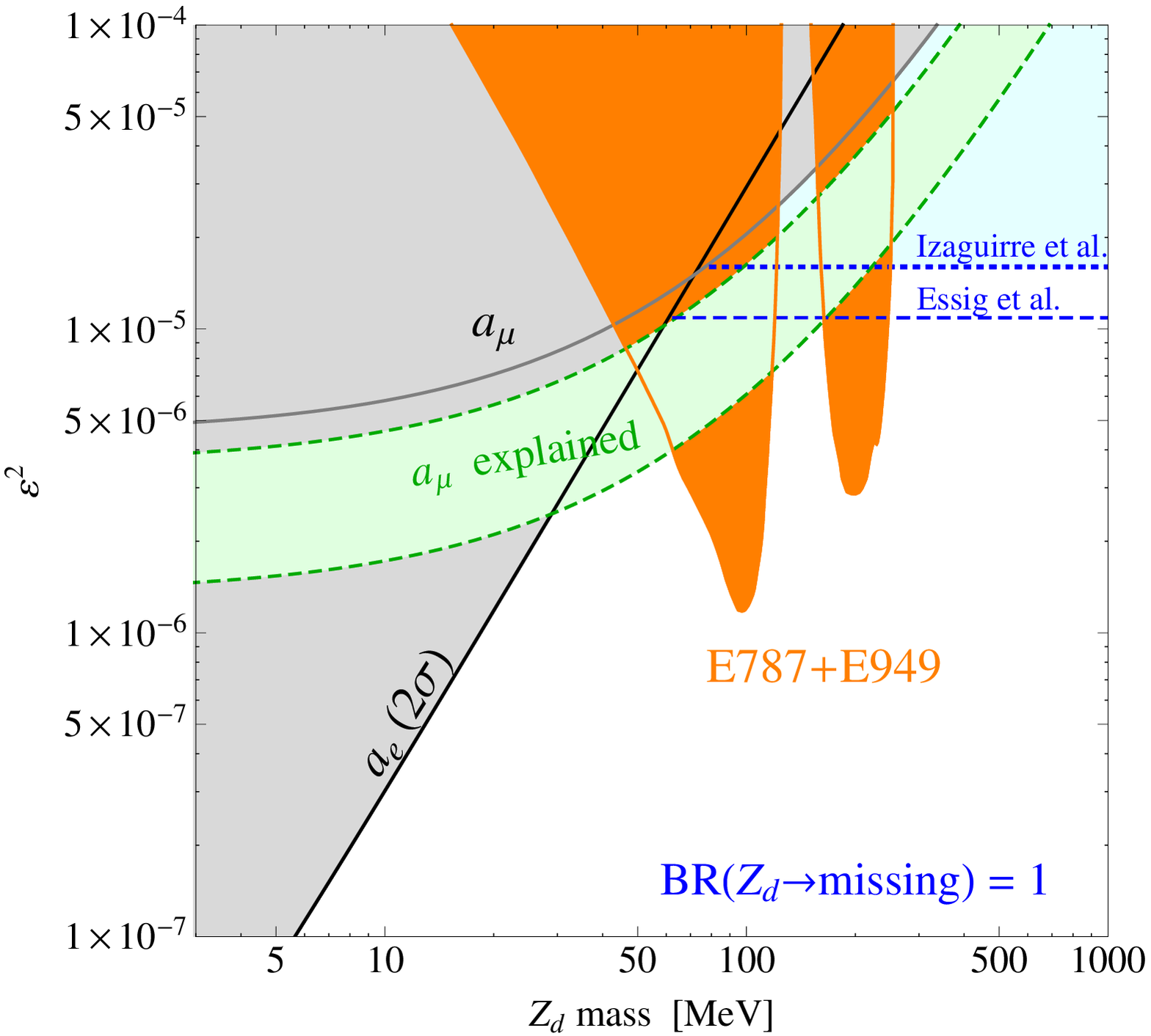}}
~~~~~
\subfigure[]{
\includegraphics[width=0.45\textwidth,clip]{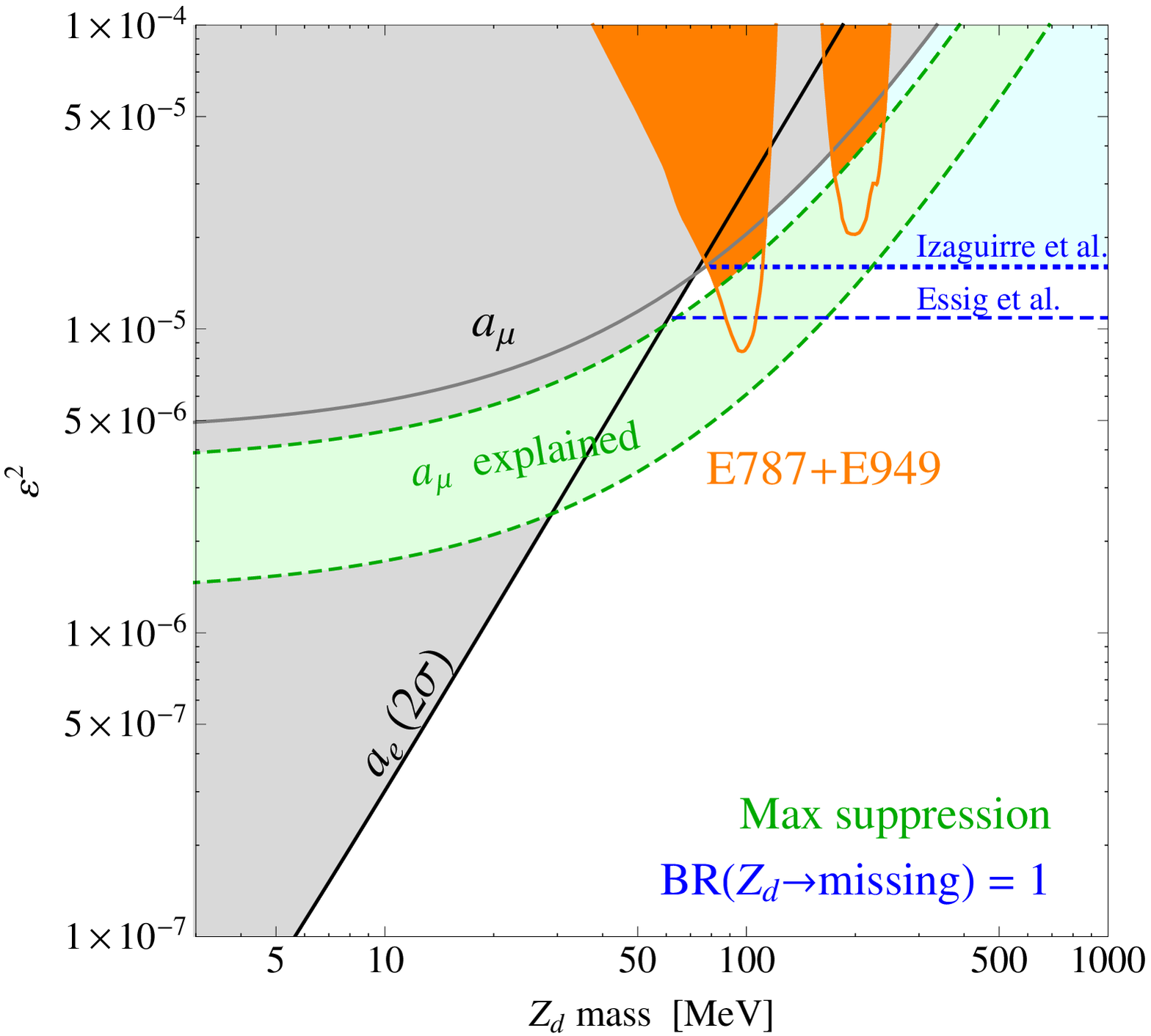}}
\end{center}
\caption{
Constraints from BNL E787+E949 experiments ($K \to \pi$ + nothing), at $95\%$ C.L., 
on the dark photon parameter space (orange area) for 
$\br(Z_d \to \text{missing}) = 1$ for (a) dark photon and (b) dark $Z$ with maximum suppression.
Also illustrated there are constraints from $e^+ e^- \to \gamma$ + `invisible' based on BaBar data as given in Ref.~\cite{Izaguirre:2013uxa} by Izaguirre et al.
and Ref.~\cite{Essig:2013vha} by Essig et al.
}
\label{fig:KaonDarkPhoton}
\end{figure*}

Figure~\ref{fig:KaonDarkPhoton} (a) shows the resulting
constraints of this ``$K \to \pi$ + nothing'' search on the dark photon model for $\br (Z_d \to \text{missing}) = 1$, 
but scaling to $95\%$ C.L., using Eq.(\ref{Kdphot}).
Rather large areas in the $a_\mu$ favored green band are excluded, i.e. the orange shaded regions around $m_{Z_d} \sim 100$, $200 ~\mev$.  The dotted and dashed  lines correspond to the constraints from $e^+ e^- \to \gamma$ + `invisible',
adapted from Refs.~\cite{Izaguirre:2013uxa} and \cite{Essig:2013vha}, respectively, based on BaBar results \cite{Aubert:2008as}.

We see that these bounds together eliminate much of the $a_\mu$ band, leaving only small regions of parameter space
to accommodate the $g_\mu-2$ discrepancy.

The CERN NA62 \cite{Ruggiero:2013oxa} and the proposed Fermilab ORKA experiments \cite{Worcester:2012rd} (precision measurements of $K \to \pi$ + nothing and other rare $K$ decays) will increase the $K \to \pi Z_d$ sensitivity by at least an order of magnitude, and the kaon decay exclusion
curves in Fig.~\ref{fig:KaonDarkPhoton} may be accordingly lowered.

\section{\boldmath Dark $Z$ and $Z - Z_d$ mass mixing}
\label{sec:darkZ}
While the dark photon, whose coupling is mainly
proportional to the electromagnetic coupling, can be realized using a relatively simple mechanism for the $Z_d$ mass
(a condensing scalar Higgs singlet or Stueckelberg mechanism \cite{Ruegg:2003ps}), a more general Higgs sector - for instance, a two Higgs doublet model (2HDM) - could
lead to mass mixing of $Z_d$ with the SM $Z$ \cite{Davoudiasl:2012ag}.  In this expanded framework,
the interaction Lagrangian of the $Z_d$ with the SM fermions includes both $\gamma - Z_d$
mixing as well as $Z - Z_d$ mixing
\beq
{\cal L}_\text{dark $Z$} = -\left(\eps e J^\mu_{em} + \eps_Z \frac{g}{2 \cos\theta_W} J^\mu_{NC} \right) {Z_d}_\mu\,, \label{eq:LdarkZ}
\eeq
where
\beq
J^\mu_{NC} \equiv (T_{3f} - 2 Q_f \sin^2\theta_W) \bar f \gamma^\mu f - T_{3f} \bar f \gamma^\mu \gamma_5 f
\eeq
is the weak neutral current and $T_{3f} = \pm 1/2$ \cite{Davoudiasl:2012ag}.
An additional parameter $\eps_Z$ is present to describe the $Z - Z_d$ mixing.
In this case, the vector state $Z_d$ is dubbed a ``dark $Z$'' to emphasize that it also has $Z$-like couplings.

In the dark $Z$ model, there are 3 independent parameters needed to describe the phenomenology:
the dark $Z$ mass $m_{Z_d}$, kinetic mixing parameter $\eps$, and the $Z - Z_d$ mass mixing parameter $\eps_Z$.
Thus, the dark photon model can be viewed as a special case (the $\eps_Z = 0$ limit) of a more general dark $Z$ model.
(We note that kinetic mixing will yield $\eps_Z \sim \eps \, m_{Z_d}^2/m_Z^2$.  We do not consider that small
effect, for $m_{Z_d} < 1$~GeV, here.)

The $Z - Z_d$ mixing gives rise to interesting phenomenological features,
such as providing a new low mass mediator of parity violation, that are
typically absent in the light dark photon models.  Also, at energies
large compared to $m_{Z_d}$, the longitudinally polarized $Z_d$ dominates and has an
enhanced coupling of order $(E/\mzd) \, \eps_Z$ (for $E\gsim m_{Z_d})$.  The Goldstone boson equivalence theorem \cite{GET} implies that
the longitudinal dark $Z$ mode at high energies exhibits properties similar to an axion and one can use computations
involving the latter to estimate the rates for processes associated with the former \cite{Davoudiasl:2012ag}.

The $Z - Z_d$ mass mixing parameter $\eps_Z$ in Eq.~\eqref{eq:LdarkZ} is further parametrized by
\beq
\eps_Z \equiv \frac{m_{Z_d}}{m_Z} \delta
\eeq
and the mass-squared matrix (in the $\eps = 0$ limit) can be written as
\beq
M_{Z Z_d}^2 \simeq \mat{m_Z^2 & -\delta \, m_Z m_{Z_d} \\ -\delta \, m_Z m_{Z_d} & m_{Z_d}^2}
\eeq
for $m_{Z_d}^2 \ll m_Z^2$.
For more details about the formalism, see Ref.~\cite{Davoudiasl:2012ag}.

The bounds on $\delta$ come from various experiments including low energy parity violation, Higgs decays, and flavor changing rare meson decays.
The typical bounds are $|\delta| \lsim 10^{-2} - 10^{-3}$, depending on the mass and decay branching ratio of $Z_d$ \cite{Davoudiasl:2012ag}.
The low momentum transfer ($Q$) parity violation experiments provide significant constraints on the parameter space, as the effect vanishes for $Q^2 \gg m_{Z_d}^2$.
They include atomic parity violation and polarized electron scattering experiments \cite{Davoudiasl:2012ag,Davoudiasl:2012qa}.

The dark $Z$ model opens a new window into the ``dark sector''
through the Higgs boson at the LHC experiments \cite{Davoudiasl:2012ag,Lee:2013fda,Davoudiasl:2013aya}.
Unlike the simple dark photon model, the
dark $Z$ leads to a small but potentially
measurable $H \to Z Z_d$ decay, which is from the SM $H \to Z Z$ process with a $Z$ replaced with $Z_d$ through $Z - Z_d$ mixing.
Because of the small $Z_d$ mass, it is an on-shell decay process producing a boosted $Z_d$ with the
aforementioned enhancement for longitudinal polarization.
The recently discovered SM-like Higgs boson \cite{ATLAS:2013oma,CMS:ril} provides a constraint on the dark $Z$ boson.
The charged Higgs boson - from a 2HDM realization of dark $Z$ - may escape current LHC searches as it can dominantly decay into $Z_d$ final states as $W Z_d$ \cite{Davoudiasl:2014mqa} or $W Z_d Z_d$ \cite{Lee:2013fda} depending on the masses of the non SM-like scalars.
For a detailed quantitative study, see Ref.~\cite{Kong:2014jwa}.

\section{\boldmath Rare Kaon decays in the presence of $Z - Z_d$ mass mixing}
\label{sec:kaonrevisit}
We now revisit the $K^\pm \to \pi^\pm Z_d$ decay in the dark $Z$ model.
This process was discussed in Ref.~\cite{Davoudiasl:2012ag} with the $Z_d$ replaced by an axion,
which is possible for the longitudinally polarized $Z_d$ (Goldstone equivalence theorem).
Here we employ a similar but more comprehensive approach.
We take a coupling adapted from the axion approximation ($\propto \eps_Z$),
along with that from kinetic mixing ($\propto \eps$) so that we can describe both interactions and their interference effects.
The formalism for $K^\pm \to \pi^\pm Z_d$ is given in the Appendix, where more details are provided.

The more general decay width for $K^\pm \to \pi^\pm Z_d$ is given by
\beq
\Gamma (K^\pm \to \pi^\pm Z_d) = \Gamma (K^\pm \to \pi^\pm Z_d)|_\eps \left |1 \pm \frac{\delta \, B}{\eps\, A m_{Z_d}\,m_Z}\right |^2
\label{eq:approxGamma}
\eeq
with $A$ and $B$ given in the Appendix.
As discussed in the Appendix, $A$ has been assumed to be real, but a $\pm$ sign arbitrariness has been included to reflect uncertainty in the long distance amplitude sign of $A$.

\begin{figure}[t]
\begin{center}
\includegraphics[height=0.27\textwidth,clip]{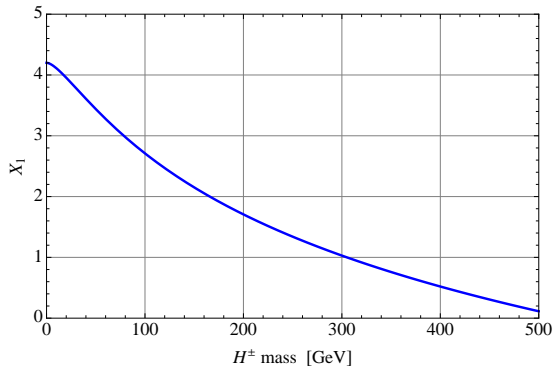}
\end{center}
\caption{$X_1$ \cite{Hall:1981bc,Freytsis:2009ct} as a function of the charged Higgs mass with $m_t = 163 ~\gev$.}
\label{fig:X1}
\end{figure}

In the pure dark $Z$ limit ($\eps = 0$), we get
\bea
&&\Gamma (K^\pm \to \pi^\pm Z_d)|_{\eps_Z} \nn \\
&=& \frac{g^6 |U_{td}^* U_{ts}|^2}{2^{20} \pi^5} (f_+)^2 X_1^2 \delta^2 \frac{m_t^4}{m_W^6 m_K^3} \sqrt{\lambda (m_K^2, m_\pi^2, m_{Z_d}^2)} \nn \\
&\times& \left[(m_K^2 - m_\pi^2)^2 - m_{Z_d}^2(2 m_K^2 + 2 m_\pi^2 - m_{Z_d}^2) \right]
\eea
where $X_1$ \cite{Hall:1981bc,Freytsis:2009ct}
depends on the charged Higgs mass and top mass and is plotted in Fig.~\ref{fig:X1}.  
Note that the above rate does not vanish as $m_{Z_d}\to 0$.
The suppression for small $m_{Z_d}$ in dark photon model does not necessarily occur in the dark $Z$ model.

The dashed curve in Fig.~\ref{fig:BRs} represents $\br (K^\pm \to \pi^\pm Z_d)|_{\eps_Z}$, with $f_+ \simeq 1$.
This plot agrees with the result in Ref.~\cite{Davoudiasl:2012ag}, which takes $m_{H^\pm} = 150 ~\gev$ and includes the small charm quark contribution, leading to a strong constraint $\delta \lsim 10^{-3}$ (except for a region near $m_{Z_d} \sim m_\pi$).

A cancellation may occur between kinetic mixing ($\eps$ term) and $Z - Z_d$ mass mixing ($\delta$ term).
In that way, the dark $Z$ may be able to evade the $K \to \pi$ + nothing search constraints.  While $A$ is 
taken to be real, $B$ is complex due to $U_{td}^* U_{ts} \simeq - (3.36 + 1.35 \, i) \times 10^{-4}$,
and there can be a cancellation between the $A$ and the real part of $B$.
A complete cancellation is impossible because Im$[B]\neq 0$,
and its contribution to $\br (K^\pm \to \pi^\pm Z_d)$ must be smaller than the experimental bound given in Ref.~\cite{Artamonov:2009sz}.

Figure~\ref{fig:KaonDarkPhoton} (b) shows the maximum suppression of the $K \to \pi$ + nothing constraint, which can be shown to be
$1.35^2/(3.36^2+1.35^2) \simeq 1/7$.  This corresponds to
\bea
\delta &=& \mp \eps \,\frac{A \,m_{Z_d}\, m_Z \text{Re}[B]}{|B|^2}
\label{max-cancel-delta} \\
&\simeq& \pm \eps \, (6.2) \, \frac{(m_{Z_d}/\text{GeV})}{X_1} \,  \, .
\eea
for a given $X_1$ value (depending on $m_{H^\pm}$).
It is interesting to observe that constraints
on $\eps$ and $\delta$ are both alleviated for the destructive interference requirement.

\section{Running of the weak mixing angle}
\label{sec:WeinbergAngle}
In this section, we consider what type of experiments can still test the dark $Z$ when the
searches for dilepton bump or missing energy (especially, $K \to \pi$ + nothing search) miss the signals due to the destructive interference effect discussed in the previous section.

The dark $Z$ can still modify neutral current phenomenology in the low $Q$ (momentum transfer) regime \cite{Davoudiasl:2012ag}.
The effective value of the weak mixing angle is
modified for $Q \lsim m_{Z_d}$, which leaves the dark $Z$ parity violating effect still visible in low energy experiments.
As suggested in Refs.~\cite{Davoudiasl:2012ag,Davoudiasl:2012qa}, low $Q^2$ polarized electron scattering
experiments in progress or proposed at various facilities (including Jefferson Lab and MAMI at Mainz) are excellent probes
of this kind of low energy new physics if $Q^2 \lsim m_{Z_d}^2$.
Another type of low energy test is
atomic parity violation \cite{Bennett:1999pd}, which requires precise understanding of the heavy atom physics.

The weak mixing angle shift by the dark $Z$ is given in Refs.~\cite{Davoudiasl:2012ag,Davoudiasl:2012qa} as
\beq
\Delta \sin^2\theta_W (Q^2) = - \eps \delta \frac{m_Z}{m_{Z_d}} \sin\theta_W \cos\theta_W f \left(Q^2 / m_{Z_d}^2\right) ,
\label{eq:DeltaWeinberg}
\eeq
with $f \left(Q^2 / m_{Z_d}^2\right) = 1 / (1 + Q^2 / m_{Z_d}^2)$ (for polarized electron scattering experiments)
and $f \left(Q^2 / m_{Z_d}^2\right) \sim 1$ (for atomic parity violation of a heavy atom \cite{Bouchiat:2004sp}).
We use $\sin^2\theta_W = 0.238$ appropriate for low energy physics.

\begin{figure*}[t]
\begin{center}
\subfigure[]{
\includegraphics[width=0.48\textwidth,clip]{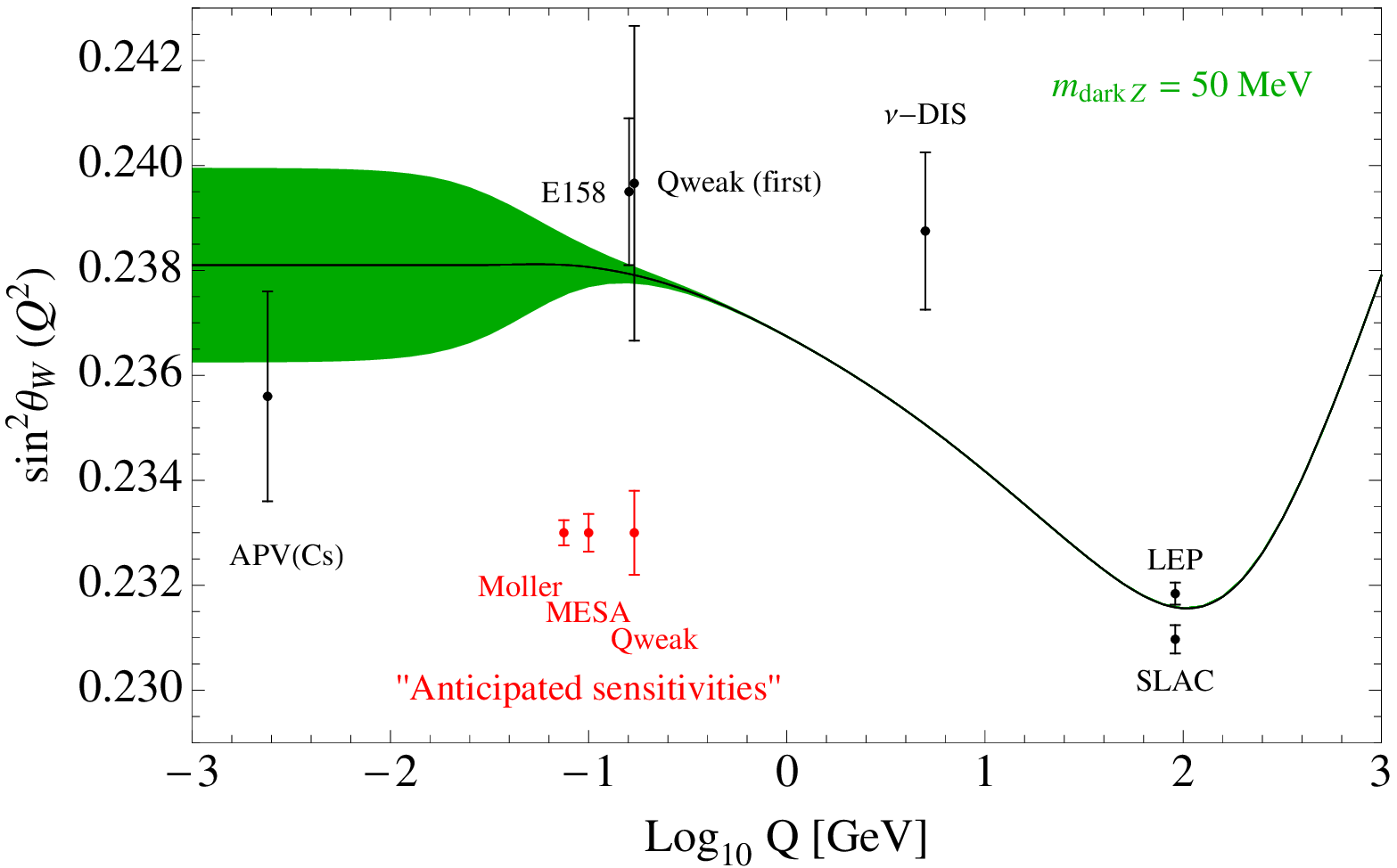}}
\subfigure[]{
\includegraphics[width=0.48\textwidth,clip]{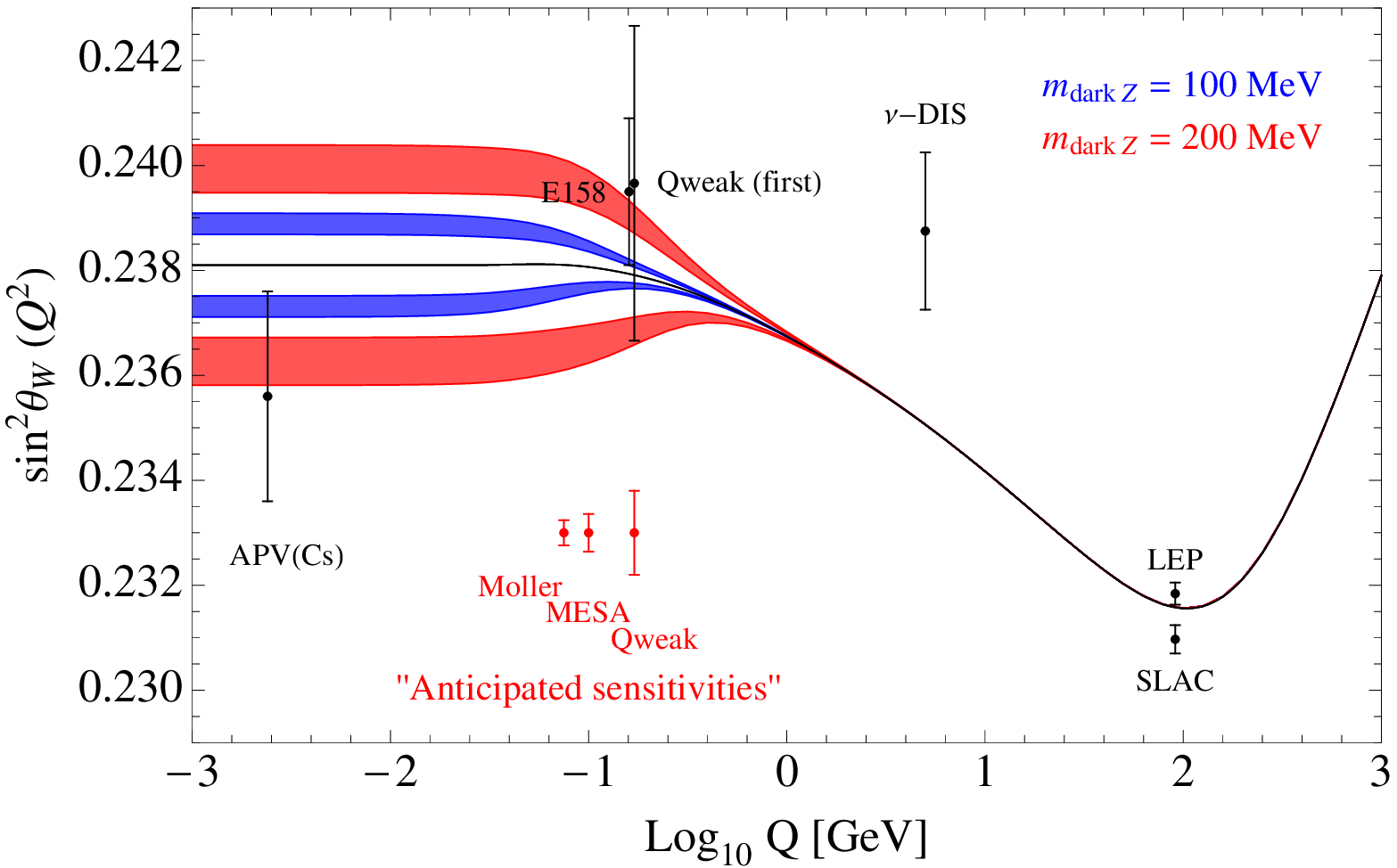}}
\end{center}
\caption{
Running of the effective weak mixing angle, $\sin^2\theta_W (Q^2)$ with energy scale $Q$.
The solid (black) curve is the SM prediction, and the shaded regions are predictions with a dark $Z$ for given masses (a) $m_{Z_d} = 50 ~\mev$ and (b) $m_{Z_d} = 100$, $200 ~\mev$ with $\eps^2$ from the $a_\mu$ green band in Fig.~\ref{fig:al}.
BNL kaon decay constraints are applied.
The red points and their error bars represent, respectively, the average $Q$ and the anticipated sensitivities for JLab Moller, Mainz MESA (P2), and JLab Qweak.
The results depend on the charged Higgs mass, and $m_{H^\pm} = 160 ~\gev$ is used.
}
\label{fig:mixingRunning}
\end{figure*}

A convenient way to illustrate the effect of quantum corrections to $\gamma - Z$ mixing is via the running of $\sin^2\theta_W (Q^2)$ introduced in Ref.~\cite{runningWeinberg}.
It describes the evolution at low energy primarily through quark loops (but with some small lepton effects) and then hits a minimum when $W^+ W^-$ loops effectively change the sign of the evolution slope at high $Q^2 > 2 m_W$.
That SM plot will be modified at low $Q^2$ by the dark $Z$ shift given in Eq.~\eqref{eq:DeltaWeinberg}.
To illustrate the effect, we give in Fig.~\ref{fig:mixingRunning} modifications of the low $Q^2$ dark $Z$ effect for specific values of $m_{Z_d} = 50 ~\mev$, $100 ~\mev$ and $200 ~\mev$.
Definite nonzero correction bands are predicted for $m_{Z_d} = 100 ~\mev$, $200 ~\mev$ assuming that the dark $Z$ solves the $g_\mu-2$ discrepancy and satisfies the bounds in Fig.~\ref{fig:KaonDarkPhoton}.

The shaded regions (potential deviation of weak mixing angle) clearly show that
the effect of the dark $Z$ is visible for $Q \lsim m_{Z_d}$.  
The two branches of each curve correspond to a potential sign ambiguity of $\eps \delta$ in Eq.~(\ref{eq:DeltaWeinberg})
that could result from QCD effects in the relative sign between $A$ and Re$[B]$.
In principle, this sign, required for the cancellation between the two amplitudes, may be determined through a detailed analysis of QCD corrections to the amplitudes.
However, such a study is outside the scope of this paper.  
We simply plot both possibilities to show the form of the expected effects in each case.

There are ongoing or planned low energy polarized electron scattering experiments including JLab Qweak
($ep$) \cite{Androic:2013rhu}, JLab Moller ($ee$) \cite{Moller}, and MESA P2 ($ep$) \cite{MESA}.
For recent reviews on the low energy weak mixing angle measurements, see Refs.~\cite{Kumar:2013yoa,Erler:2014fqa}.
As is clear from the sensitivities indicated in Fig.~\ref{fig:mixingRunning}, a deviation in the weak mixing angle by dark $Z$ can be large enough to be tested by these
low energy parity measurements.  Furthermore, using the difference in average momentum
transfer ($Q$) of these experiments, including the atomic parity violation, it may be
possible to fit the data to constrain the mass and couplings of $Z_d$ if deviations from the SM predictions arise.

\section{Summary and Conclusions}
\label{sec:conclusion}
In this paper, we examined the properties of a hypothetical dark photon (or dark $Z$),
which was proposed to address various astrophysical anomalies and the deviation in the muon anomalous magnetic moment.
We discussed current bounds and near future sensitivity, including a detailed discussion 
of implications from the electron anomalous magnetic moment.

We also considered scenarios in which the dark photon decays primarily into light dark matter or other invisible particles, where the typical searches assuming $\br(Z_d \to \ell^+ \ell^-) = 1$ would not be
sensitive to the signal.
This case is timely as the current experiments based on bump searches in
addition to the electron and muon anomalous magnetic moment are tending to exclude most of the
preferred $\eps$ parameter region that can explain the $3.6 \sigma$ muon $g-2$ discrepancy.
Considering that there are active analysis of existing data and numerous future experiments,
it is expected the whole region will be tested soon and possibly ruled out.
Of course, a more interesting outcome would be discovery of the dark sector.

Interestingly, the $K \to \pi$ + nothing searches (BNL E787+E949) can exclude the scenario of dominant $Z_d$ decay into invisible particles in large parts of the dark photon parameter space.  Used in conjunction with recent bounds from $e^+ e^- \to \gamma$ + `missing energy' \cite{Izaguirre:2013uxa,Essig:2013vha} (based on BaBar results \cite{Aubert:2008as}), one can significantly constrain the
$g_\mu-2$ preferred $\eps$ parameter space.
We emphasized that for the dark $Z$, which is essentially a dark photon with a more general coupling,
we can potentially evade the current rare kaon decay constraints on 
missing energy searches due to the possibility of a cancellation 
between the kinetic mixing and $Z - Z_d$ mass mixing.
As the light $Z_d$ contribution to the muon anomalous magnetic moment is independent of
its decay branching ratio, the $Z_d$ can still remain as the solution to the muon anomaly.
In this case, low energy polarized electron scattering as well as atomic parity violation
predictions can provide sensitive tests of that scenario.

Acknowledgments:
We would like to thank R. Essig for discussions.
This work was supported in part by the United States DOE under Grant No.~DE-AC02-98CH10886, No.~DE-AC05-06OR23177, and by the NSF under Grant No.~PHY-1068008.
W.M. acknowledges partial support as a Fellow in the Gutenberg Research College.
H.L. appreciates hospitality during his visit to BNL.

\appendix
\section{Formalism}

The amplitude for $K^\pm (k) \to \pi^\pm (p) + Z_d (q)$ is given by
\bea
&&{\cal M} (K^\pm \to \pi^\pm Z_d) \nn \\
&=& \left( \eps A \left< \pi^\pm(p) | q^2 \bar s_L \gamma_\mu d_L - q_\mu q^\nu \bar s_L \gamma_\nu d_L | K^\pm(k) \right> \right. \nn \\
&& \left. \pm \delta \frac{m_{Z_d}}{m_Z} B \left< \pi^\pm(p) | \bar d_L \gamma_\mu s_L | K^\pm(k) \right>\right) \epsilon^{*\mu} (q) \\
&=& \frac{1}{2} f_+ (q^2) \left( \eps m_{Z_d}^2 A \pm \delta \frac{m_{Z_d}}{m_Z} B \right) (k+p)_\mu \, \epsilon^{*\mu} (q) ~~~~
\eea
where we have used $\epsilon^\mu (q) \, q_\mu = 0$ 
and the hadronic matrix elements 
\bea
\left< \pi^\pm(p) | \bar s \gamma_\mu d | K^\pm(k) \right> &=& f_+ (q^2) \, (k+p)_\mu \, , \\
\left< \pi^\pm(p) | \bar s \gamma_\mu \gamma_5 d | K^\pm(k) \right> &=& 0 \, ,
\eea
with $f_+ (0) = 1$.
We have allowed for a $\pm$ arbitrariness in the relative sign of $A$ and $B$ because $A$ is dependent on long distance QCD effects that could change its sign.
We also assume that $A$ is real in our discussion.
In principle, it could have an imaginary part.
We avoid that issue by focusing on $m_{Z_d} < 2 m_\pi$, since imaginary parts are due primarily to 2 pion intermediate state in the chiral expansion \cite{D'Ambrosio:1998yj}.
Taking the formalism introduced in Ref.~\cite{Pospelov:2008zw} 
(for $A$) and Refs.~\cite{Hall:1981bc,Freytsis:2009ct} (for $B$), we have
\bea
A &=& \frac{1}{(4\pi)^2} \frac{e W}{m_K^2 (f_+ / 2)} \label{eq:A2} \label{eq:AP} \\
B &=& \frac{1}{(4\pi)^2} \frac{g^3 m_t^2 m_Z}{8 m_W^3} \left( U_{td}^* U_{ts} \right) X_1 \label{eq:B2}
\eea
where we have included only a dominant top quark loop term in $B$.
(For an approach based on the SM loop induced photon and $Z$ couplings, see Ref.~\cite{Inami:1980fz}.)
The dark photon case corresponds to $\delta = 0$, and the pure dark $Z$ limit is obtained for $\eps = 0$.  
The function $W$ is given in Ref.~\cite{D'Ambrosio:1998yj}.  
It was approximated by $W^2 \approx 3 \times 10^{-12} \, (1 + 2 q^2 / m_K^2)$ \cite{Pospelov:2008zw,Essig:2013vha}.
For $W / f_+$, we use $\pm 1.73 \times 10^{-6}$ where a sign arbitrariness is allowed.
The function $X_1$ \cite{Hall:1981bc,Freytsis:2009ct}, plotted in Fig.~\ref{fig:X1},
depends on the charged Higgs mass and top mass, for which we use $m_t = 163 ~\gev$
(the QCD corrected value in the $\overline{\text{MS}}$ scheme).

The decay width for $K^\pm \to \pi^\pm + Z_d$ is then
\beq
\Gamma (K^\pm \to \pi^\pm Z_d) = 4\pi \frac{\sqrt{\lambda (m_K^2, m_\pi^2, m_{Z_d}^2)}}{64 \pi^2 m_K^3} \sum_\text{pol} | {\cal M} |^2 \label{eq:fullGamma}
\eeq
with $\lambda(x, y, z) \equiv x^2 + y^2 + z^2 - 2xy - 2yz - 2zx$ and the amplitude squared written as
\bea
&&\sum_\text{pol} |{\cal M}|^2 \nn \\
&=& \frac{1}{4} (f_+)^2 \left[ \left(\frac{m_K^2 - m_\pi^2}{m_{Z_d}}\right)^2 - (2 m_K^2 + 2 m_\pi^2 - m_{Z_d}^2) \right] \nn \\
&& \times \left| \eps m_{Z_d}^2 A \pm \delta \frac{m_{Z_d}}{m_Z} B \right|^2 .\label{eq:amplitudeSqr}
\eea



\end{document}